\documentclass[a4paper]{article}

\usepackage{INTERSPEECH2019}
\usepackage{graphicx}
\usepackage{subcaption}
\usepackage{xcolor}

\title{Sub-band Convolutional Neural Networks \\for Small-footprint Spoken Term Classification}
\name{Chieh-Chi Kao, Ming Sun, Yixin Gao, Shiv Vitaladevuni, Chao Wang}
\address{Alexa Speech, Amazon}
\email{\{chiehchi,mingsun,yixigao,shivnaga,wngcha\}@amazon.com}

\begin{document}

\maketitle
\begin{abstract}
This paper proposes a Sub-band Convolutional Neural Network for spoken term classification. 
Convolutional neural networks (CNNs) have proven to be very effective in acoustic applications such as spoken term classification, keyword spotting, speaker identification, acoustic event detection, etc.
Unlike applications in computer vision, the spatial invariance property of 2D convolutional kernels does not fit acoustic applications well since the meaning of a specific 2D kernel varies a lot along the feature axis in an input feature map. 
We propose a sub-band CNN architecture to apply different convolutional kernels on each feature sub-band, which makes the overall computation more efficient.
Experimental results show that the computational efficiency brought by sub-band CNN is more beneficial for small-footprint models.
Compared to a baseline full band CNN for spoken term classification on a publicly available Speech Commands dataset, the proposed sub-band CNN architecture reduces the computation by 39.7\% on commands classification, and 49.3\% on digits classification with accuracy maintained.
\end{abstract}


\section{Introduction}

With the rapid development of public available datasets (e.g. spoken term classification~\cite{Warden_GoogleSpeechCommand}, speaker identification~\cite{Nagrani17,Chung18b}, acoustic event classification/detection~\cite{AudioSet,DCASE2017challenge}, etc.), state-of-the-art models for various acoustic applications can be trained with a large amount of annotated data. 
CNN-based architectures have achieved state-of-the-art performance in keyword spotting~\cite{Sainath15}, speech recognition~\cite{AbdelHamid2012CNN_SR,Qian16}, speaker identification~\cite{Nagrani17,Chung18b}, acoustic event classification~\cite{Takahashi2016DeepCN,Hershey2017,Shi2019,Tang2019}, and rare acoustic event detection~\cite{Lim2017,Kao18}.
CNNs have shown performance superior to feed-forward Deep Neural Networks (DNNs) in various acoustic applications due to the following reasons. 
First, DNN is not good at modeling the strong correlations in time and frequency of acoustic signal.
Second, DNNs are not able to model shift of formants in speech signals. 
Instead, CNNs are able to capture local patterns that model the correlations properly and detect shift of formants by sharing the weights of 2D kernels (\textit{time}$\times$\textit{feature}) across different locations in the input feature space.

An important property of CNN is shift invariance (also known as spatial invariance), which allows CNN to detect patterns even if it does not appear at exactly the same location as samples seen in the training set.
Weights of learned 2D convolutional kernels are shared across different locations in the input feature map. 
This property is desirable for visual detection tasks, where the physical meaning of two dimensions in the input feature map are the same (i.e. $x$-axis and $y$-axis in an image).
However, the spatial invariance property of 2D kernels does not fit acoustic applications well since the physical meaning of a specific 2D kernel holds only along the time axis in a input feature map, but varies a lot along the feature axis. 
To keep the shift invariance property locally, we propose an architecture of sub-band CNN by limiting the weight sharing within a certain sub-band on the feature axis.

\begin{figure*}[t!]
    \centering
    \begin{subfigure}[b]{0.28\textwidth}
        \includegraphics[width=\textwidth]{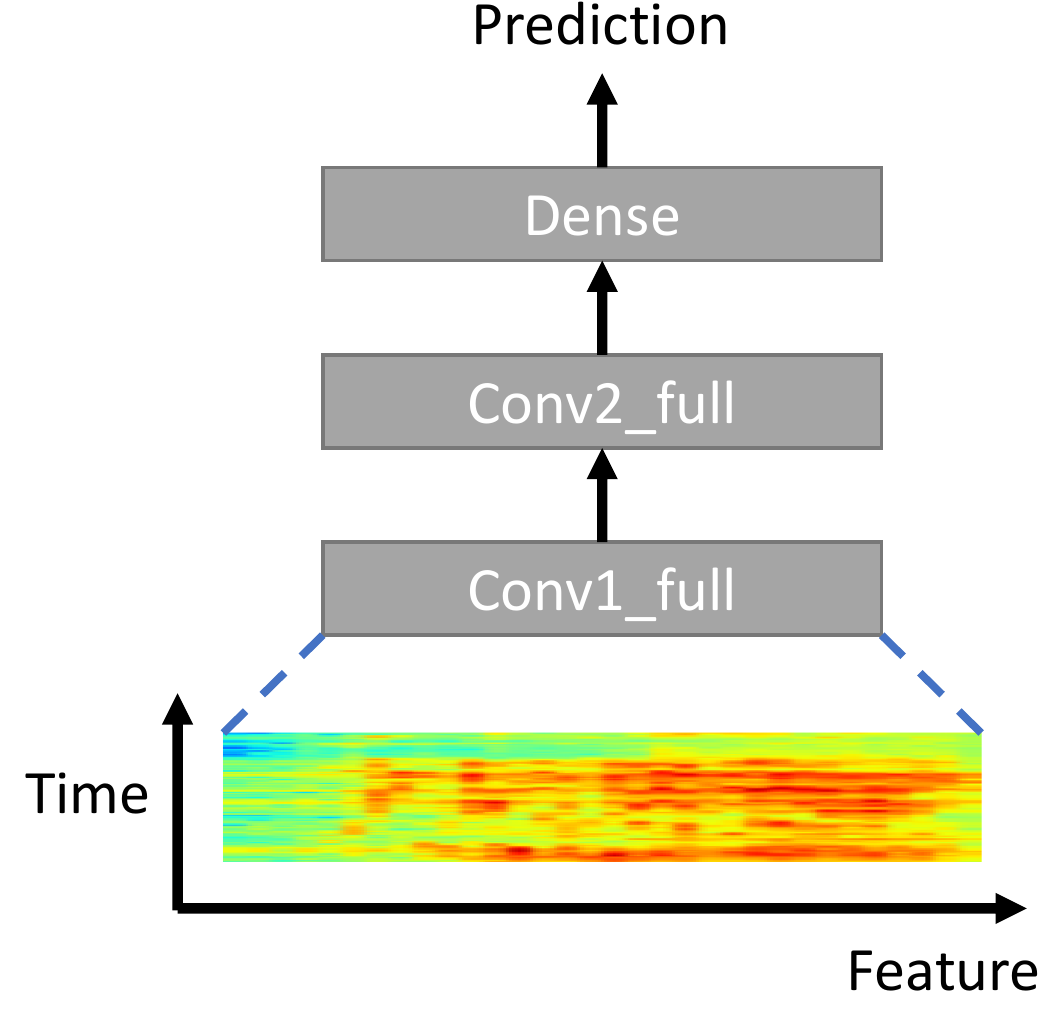}
        \caption{Full band (baseline)}
        \label{fig:fullband}
    \end{subfigure}
    ~ 
    \begin{subfigure}[b]{0.33\textwidth}
        \includegraphics[width=\textwidth]{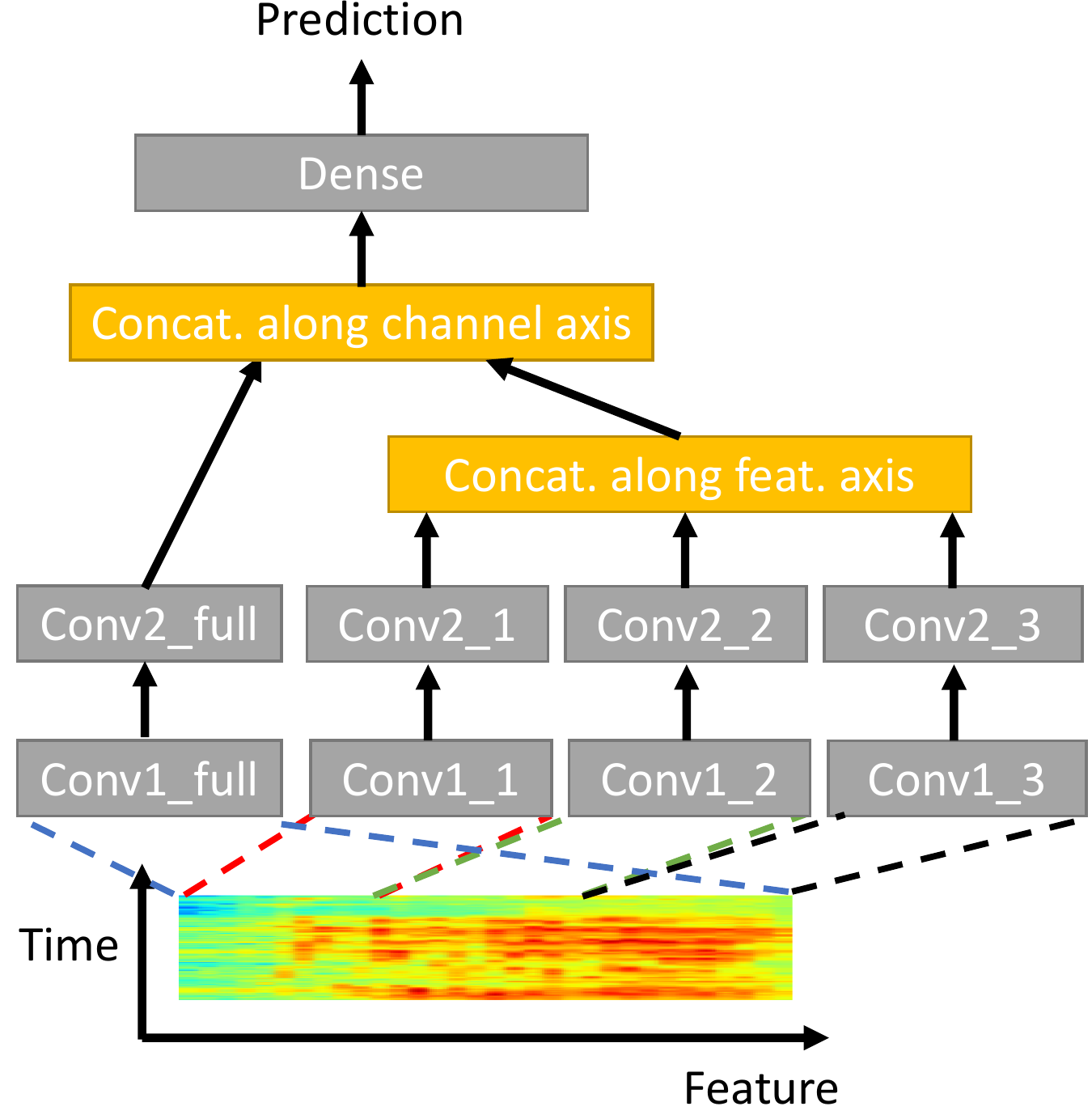}
        \caption{Full band plus non-overlapped sub-band}
        \label{fig:multiband}
    \end{subfigure}
    ~ 
    \begin{subfigure}[b]{0.28\textwidth}
        \includegraphics[width=\textwidth]{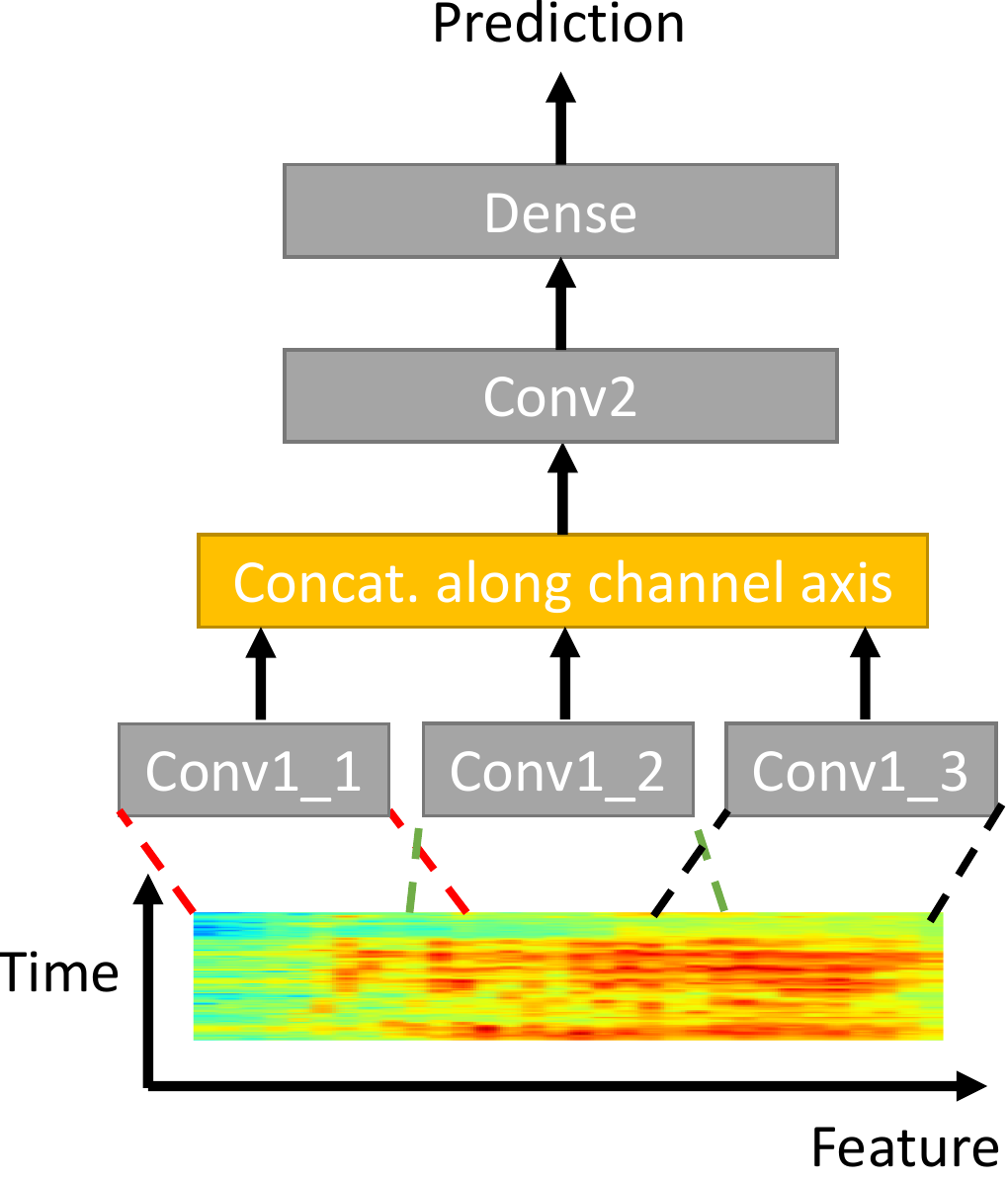}
        \caption{Overlapped sub-band (proposed)}
        \label{fig:subband}
    \end{subfigure}
    \caption{CNN models with different weight sharing methods. (a) The baseline model proposed in~\cite{Sainath15}. (b) Applying the multi-band approach proposed in~\cite{Takahashi2017} to the baseline model. (c) The proposed overlapped sub-band CNN. For the easiness of illustration, the $x$-axis is set as feature in this figure, which is different from conventional settings.}\label{fig:architectures}
    \vspace{-0.5cm}
\end{figure*}

In this work, we experimented the proposed architecture on Google Speech Commands dataset~\cite{Warden_GoogleSpeechCommand}, which provides a common benchmark for keyword spotting and spoken term classification. 
The goal of both tasks is to detect a relatively small set of predefined keywords in an utterance.
Different from keyword spotting that has been widely used for virtual assistants (e.g. Amazon Alexa, Google Assistant, Apple Siri), spoken term classification does not have the low-latency constraint since the classification is done at utterance level.
Previous works~\cite{Chen2014,He2017,Arik2017,Tang2018} showed that neural networks are very effective in keyword spotting.
As tremendous efforts are dedicated into the discovery of effective CNN architectures for further advancing the performance, we argue that it is also important to investigate into effective ways for utilizing computational resource at inference time.
Since most of the applications mentioned above run on mobile devices or smart speakers, a model with small memory footprint and low computational budget is required.
While previous works used low-rank SVD~\cite{Tucker2016,Sun2017} and knowledge distillation~\cite{Lu2017,Pang2018,Shi2019IS} to make neural networks more compact, this work focuses on how to utilize the computational resource efficiently for CNNs.
Compared with residual network for small-footprint keyword spotting (5.65M multiplications)~\cite{Tang2018}, we explored a regime with much lower computational resource for spoken term classification, which is more than 20${\times}$ reduction in the number of multiplications. 
\footnote{We use FLOPS as the measurement of computation complexity in this paper, while the number of multiplication is used in~\cite{Tang2018}. For the proposed model, about half of the FLOPS are multiplications. We use this rate to convert FLOPS to the number of multiplication.}


We propose a simple approach to utilize the computational resource efficiently for CNNs that are dedicated for acoustic applications. 
Our approach applies different sets of convolutional kernels at each feature sub-band.
Feature maps extracted from each sub-band are concatenated along the channel axis, and then fed into the next convolutional layer.
Limited weight sharing (LWS) for convolutional layers has been proven very effective in speech recognition~\cite{AbdelHamid2012CNN_SR,AbdelHamid2013ExploringCNN}.
LWS is explored for models with a single convolutional layer in these two works.
A CNN that consists of one pair of convolution and max-pooling layers, and two fully connected hidden layers is used in ~\cite{AbdelHamid2012CNN_SR}.
LWS in either time or feature axis for models with a single layer convolutional layer are discussed in~\cite{AbdelHamid2013ExploringCNN}.
Although multi-layers CNN has been tested in~\cite{AbdelHamid2013ExploringCNN}, full weight sharing was used so that more than one convolutional layers can be stacked.
Different from these works, we apply LWS to CNNs with multiple convolutional layers, and explore different ways to concatenate features extracted from each sub-band.

Most similar to our approach of using sub-band CNN with multiple convolutional layers is the work of~\cite{Takahashi2017}, which combines multiple non-overlapped sub-band sub-networks with a full band sub-network for audio source separation.
Different from~\cite{Takahashi2017}, we use overlapped sub-band networks without an extra full band sub-network.
Overlapping between sub-bands helps to avoid information loss at the boundary between sub-bands.
\cite{Takahashi2017} used a full band sub-network to avoid this information loss brought by non-overlapped sub-bands.
However, it may introduce redundant information as well as excessive computational costs.
Our experimental results show that the proposed overlapped sub-band CNN performs better than the multi-band architecture proposed in~\cite{Takahashi2017} on spoken term classification.
Recently, Phaye et. al~\cite{Phaye2019} proposed an architecture using sub-spectrogram based CNN for acoustic scene classification. 
Different from these two works~\cite{Takahashi2017,Phaye2019} of applying LWS to CNNs with multiple convolutional layers, this work is the first one to concatenate features within a CNN rather than concatenating the feature extracted from the top most layer of a CNN.
Experimental results in Sec.~\ref{sec:concat} show that concatenating features along channel axis within CNN outperforms other concatenation methods for sub-band CNN.

In the rest of this paper, we discuss our sub-band CNN approach in Section 2, demonstrate them on two tasks (commands and digits classification) of Speech Commands dataset~\cite{Warden_GoogleSpeechCommand} in Section 3, and provide conclusion remarks in Section 4.

\begin{table}
\begin{center}
    \caption[Table caption text]{Detailed architecture of full band CNN (Fig.~\ref{fig:fullband}).}
    \label{table:arch_baseline}
    \begin{tabular}{| l | l | }
    \hline
    Layer & Full band \\ \hline
    Conv 1 (t{$\times$}f, ch, stride) &20$\times$8, $K$, 1$\times$1 \\ \hline
    Activation 1 &ReLU\\\hline
    Dropout 1 &$P$\\\hline
    Maxpool (t{$\times$}f, stride) &2$\times$2, 2$\times$2\\\hline
    Conv 2 (t{$\times$}f, ch, stride) &10$\times$4, $K$, 1$\times$1 \\ \hline
    Activation 2 &ReLU\\\hline
    Dropout 2 &$P$\\\hline
    Dense (\# outputs)&12\\
    \hline
    \end{tabular}
    \vspace{-0.8cm}
\end{center}
\end{table}

\begin{table}
\begin{center}
    \caption[Table caption text]{Detailed architecture of full band plus non-overlapped sub-band CNN (Fig.~\ref{fig:multiband}).}
    \label{table:arch_multiband}
    \begin{tabular}{| p{1.8cm} | l | l | l | l |}
    \hline
    Layer & Band 1 & Band 2 & Band 3 & Full band \\ \hline
    Conv 1 &20$\times$8,&20$\times$8, &20$\times$8, &20$\times$8,\\
    (t{$\times$}f, ch, str.) & $K$,1$\times$1 & $K$,1$\times$1 & $K$,1$\times$1 & $K$,1$\times$1 \\\hline
    Activ. 1 &ReLU&ReLU&ReLU&ReLU\\\hline
    Dropout 1 &$P$&$P$&$P$&$P$\\\hline
    Maxpool &2$\times$2,&2$\times$2,&2$\times$2,&2$\times$2,\\
    (t{$\times$}f, str.)  &2$\times$2&2$\times$2&2$\times$2&2$\times$2\\\hline
    Conv 2 &10$\times$4,&10$\times$4, &10$\times$4, &10$\times$4,\\
    (t{$\times$}f, ch, str.) & $K$,1$\times$1 & $K$,1$\times$1 & $K$,1$\times$1 & $K$,1$\times$1 \\\hline
    Activ. 2 &ReLU&ReLU&ReLU&ReLU\\\hline
    Dropout 2 &$P$&$P$&$P$&$P$\\\hline
    Concat. (axis)&\multicolumn{3}{|c|}{Feature} & -\\\hline
    Concat. (axis)&\multicolumn{4}{|c|}{Channel}\\\hline
    Dense (\# o/p)&\multicolumn{4}{|c|}{12}\\
    \hline
    \end{tabular}
    \vspace{-0.8cm}
\end{center}
\end{table}

\begin{figure*}[t!]
    \centering
    \begin{subfigure}[b]{0.48\textwidth}
        \includegraphics[width=\textwidth]{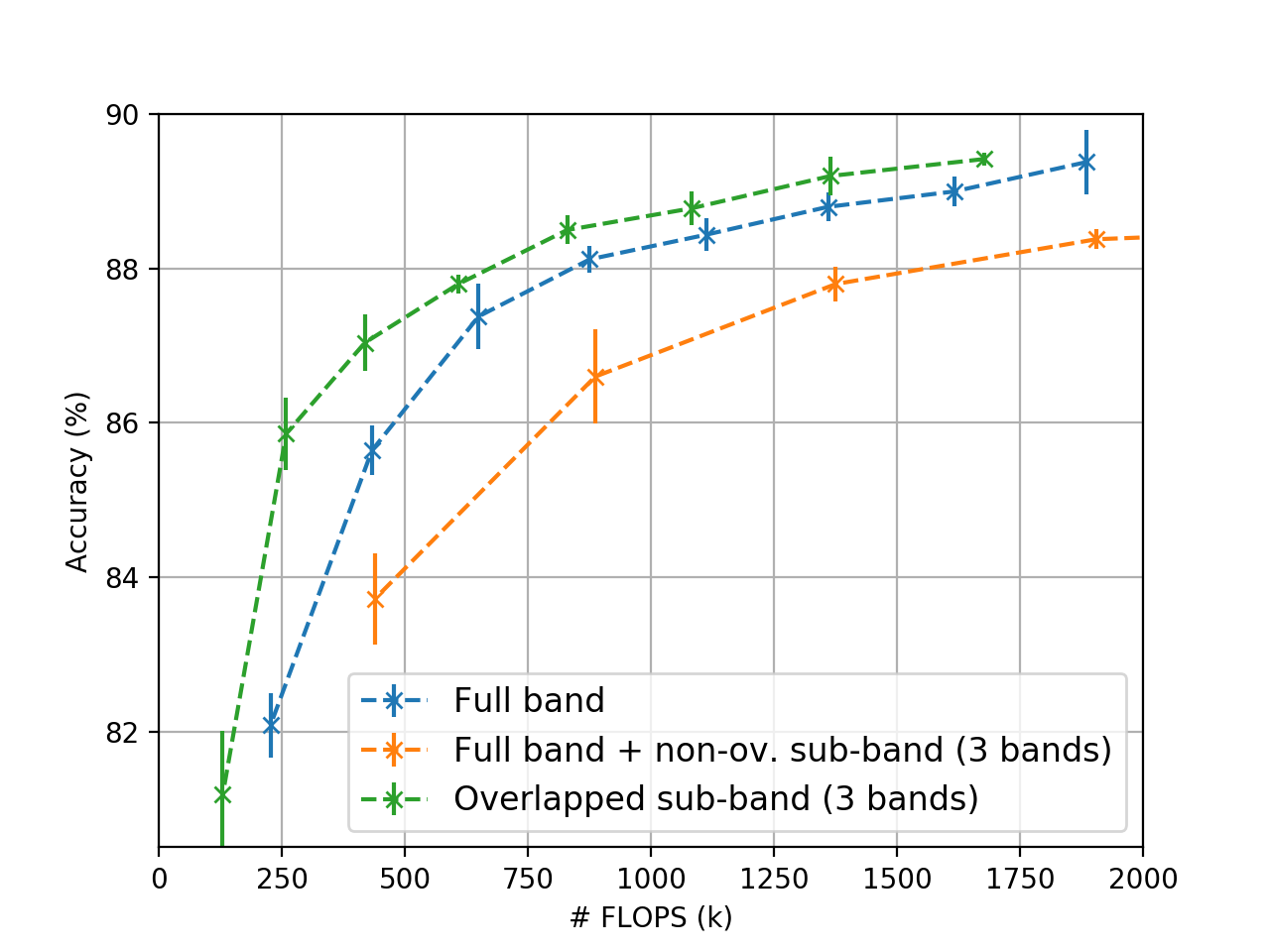}
        \caption{Commands classification}
        \label{fig:commands.main}
    \end{subfigure}
    ~ 
    \begin{subfigure}[b]{0.48\textwidth}
        \includegraphics[width=\textwidth]{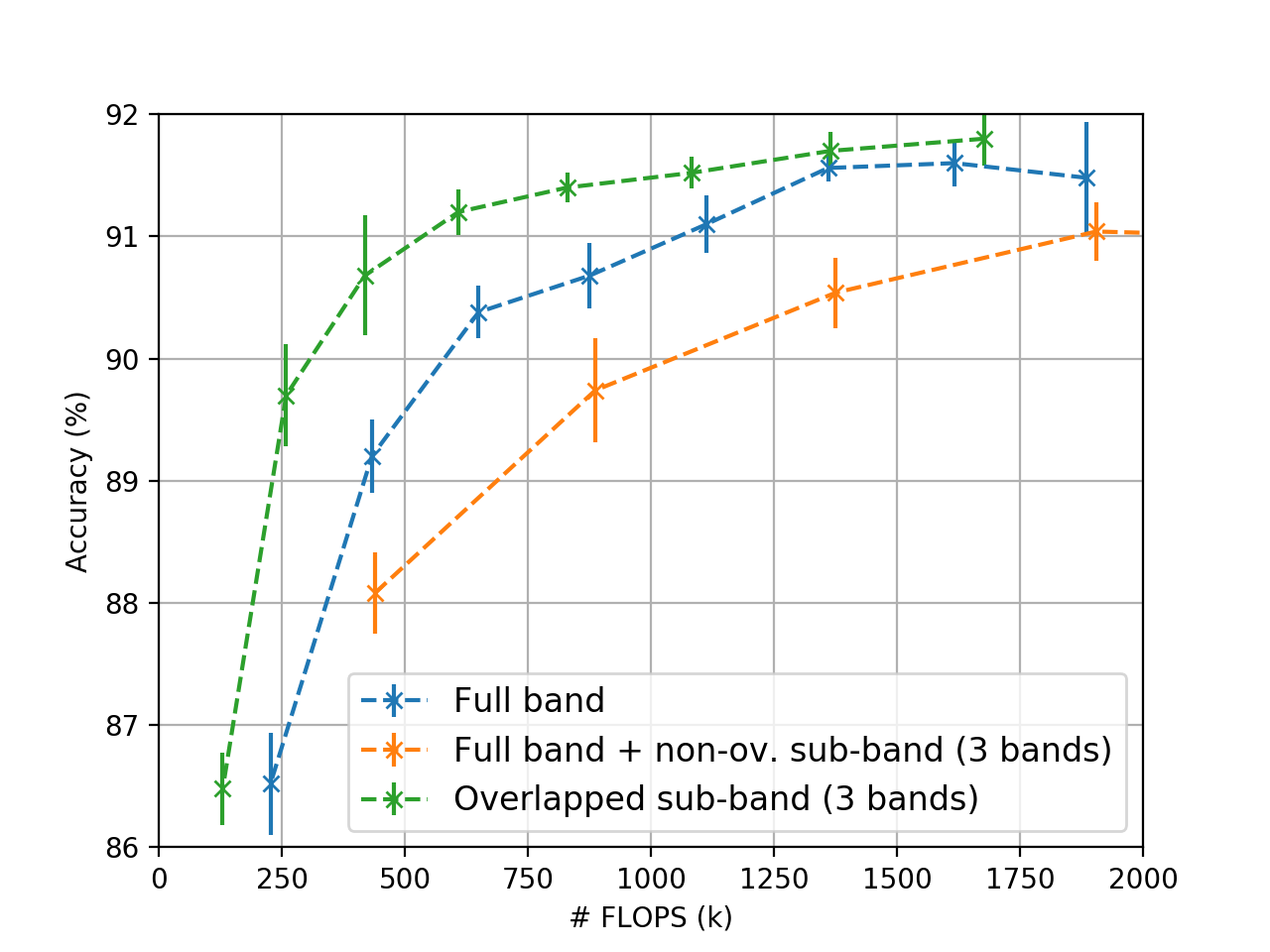}
        \caption{Digits classification}
        \label{fig:digits.main}
    \end{subfigure}
    \vspace{-0.2cm}
    \caption{Accuracy curve of different weight sharing methods on subsets of Google Speech Commands dataset~\cite{Warden_GoogleSpeechCommand}. Each data point represents an average of five trials, and the error bar is the sample standard deviation of five trials.}
    \vspace{-0.5cm}
    \label{fig:acc.main}
\end{figure*}

\section{Sub-band CNN}
We show implementation details of the proposed sub-band CNN in this section.
We chose the ``\texttt{cnn-trad-fpool3}" model proposed in~\cite{Sainath15} as our baseline model.
We used the implementation of ``\texttt{cnn-trad-fpool3}" in Tensorflow official package~\cite{tensorflow2015-whitepaper} as the baseline, which is slightly different from the original model described in ~\cite{Sainath15}.
As shown in Fig.~\ref{fig:fullband}, it consists of two convolutional layers followed by a dense layer.
The detailed architecture of the baseline model is shown in Table~\ref{table:arch_baseline}.
The convolutional layers are applied to the full band input feature map, which is equivalent to full weight sharing mentioned earlier.

We applied the proposed sub-band CNN idea to the baseline model, and Fig.~\ref{fig:subband} shows the architecture of the resulting model.
First, the input feature map is split into $B$ overlapped sub-bands ($B$=3 in Fig.~\ref{fig:subband}), and each sub-band has its own set of kernels at the first convolutional layer. 
The feature extracted from each sub-band after the first convolutional layer are concatenated along the channel axis, and then fed into the second convolutional layer.
The high-level feature extracted by the second convolutional layer is then fed into a dense layer to generate the final prediction.
Note that we set the number of kernels ($K$) in each convolutional block to be the same (i.e. there are $K$ kernels in `\texttt{Conv1\_1}', `\texttt{Conv1\_2}', ..., `\texttt{Conv1\_B}', and `\texttt{Conv2}' respectively).
The detailed architecture is shown in Table~\ref{table:arch_subband}.

\begin{figure*}[t!]
    \centering
    \begin{subfigure}[b]{0.48\textwidth}
        \includegraphics[width=\textwidth]{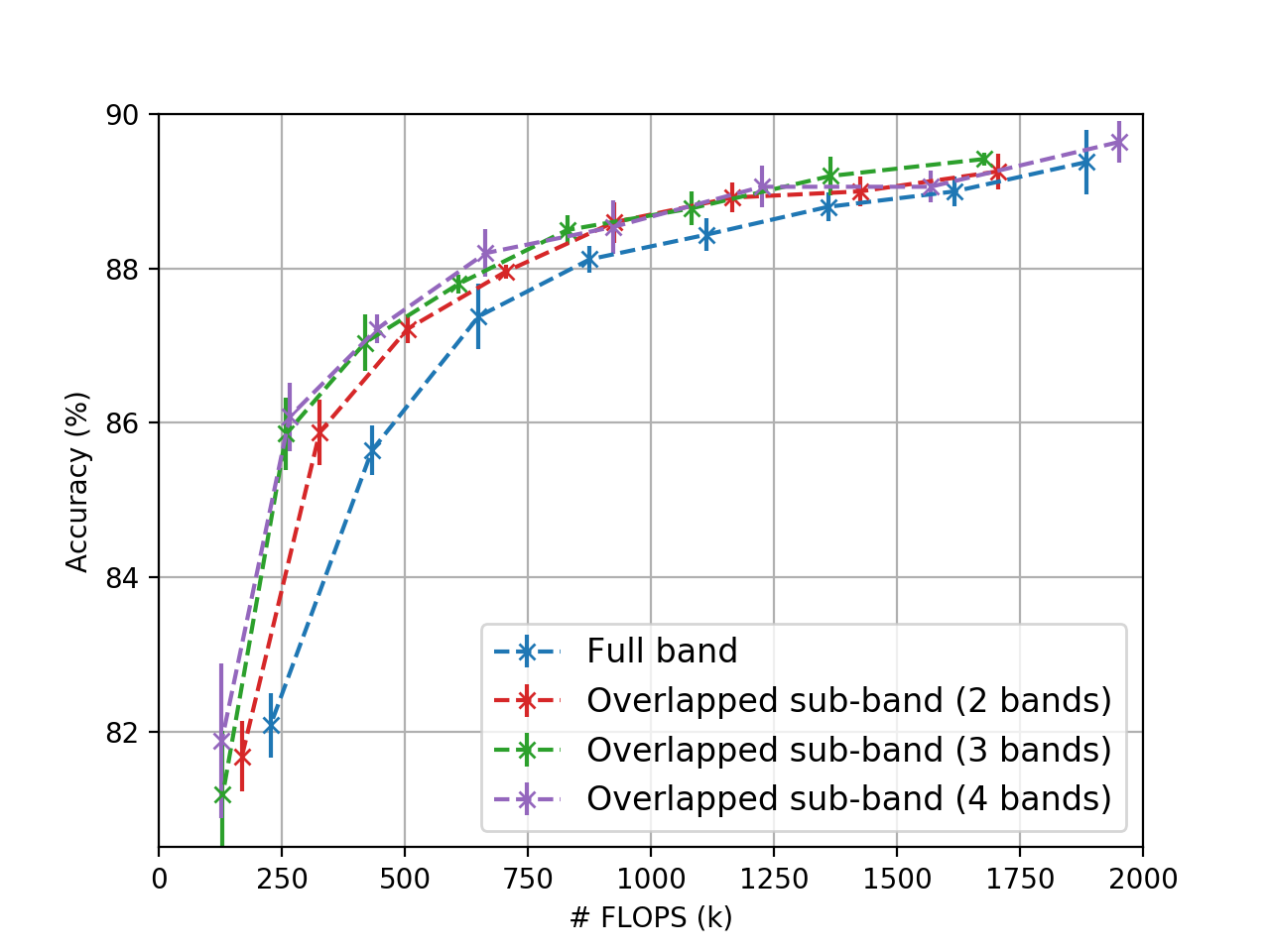}
        \caption{Different number of bands}
        \label{fig:commands.layers}
    \end{subfigure}
    ~ 
    \begin{subfigure}[b]{0.48\textwidth}
        \includegraphics[width=\textwidth]{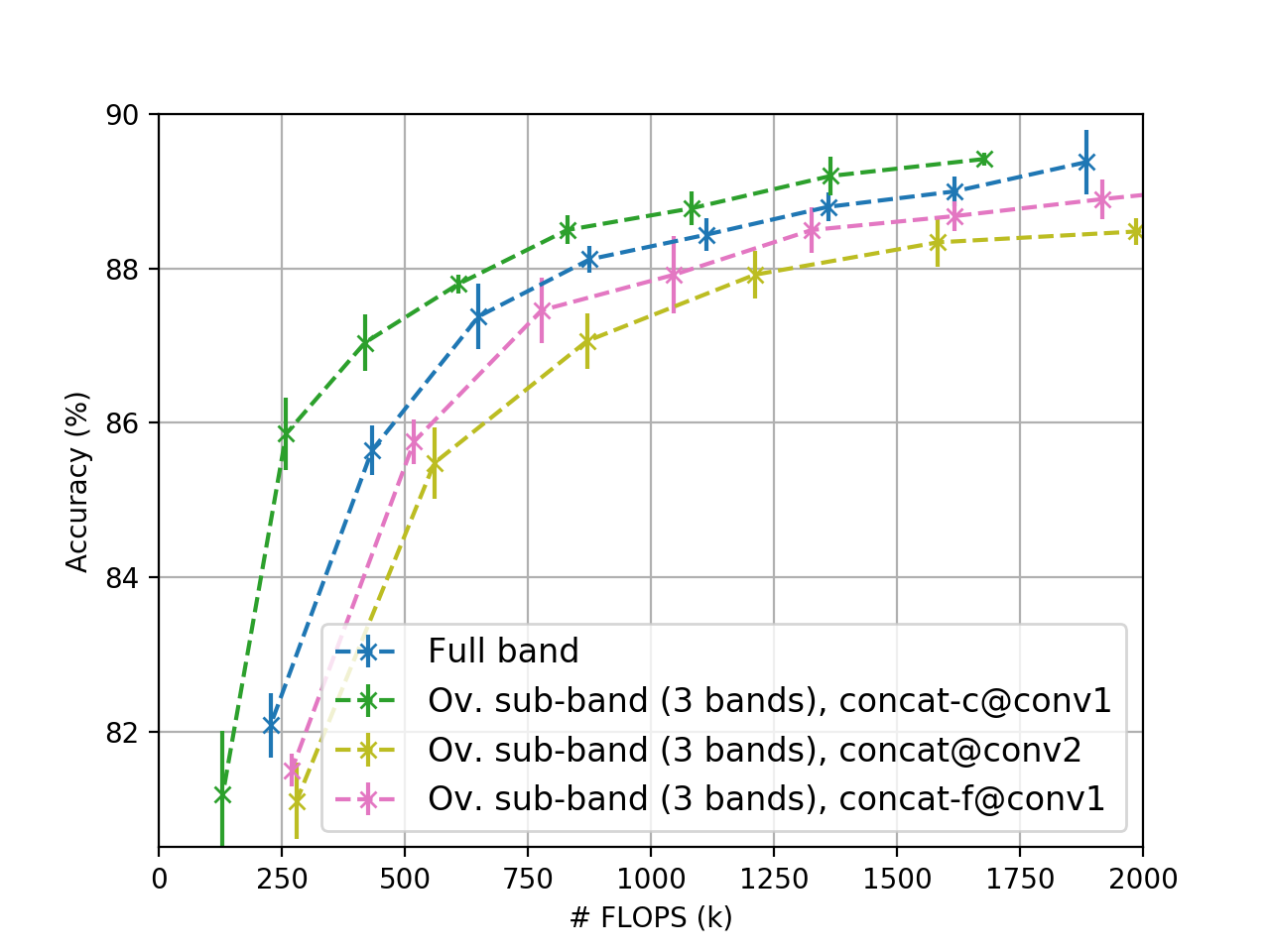}
        \caption{Different concatenation methods for sub-band features}
        \label{fig:commands.concat}
    \end{subfigure}
    \vspace{-0.2cm}
    \caption{Accuracy curve of experiments on number of sub-bands and concatenation methods for sub-band features. Commands classification is used as the testbed. Each data point represents an average of five trials, and the error bar is the sample standard deviation of five trials.}
    \vspace{-0.1cm}
    \label{fig:acc.ablation}
\end{figure*}

We experimented with different number of sub-bands (\{2,3,4\}) on commands classification task. 
As shown in Fig.~\ref{fig:commands.layers}, the model with 3 bands is comparable to the one with 4 bands, and both of them outperform the model with 2 bands.
We chose the model with 3 bands for further experiments per Occam's razor.
For comparison, we applied the multi-band approach proposed in~\cite{Takahashi2017} to the baseline model.
It is shown in Fig.~\ref{fig:multiband}, and the detailed architecture is shown in Table~\ref{table:arch_multiband}.

\begin{table}
\begin{center}
    \caption[Table caption text]{Detailed architecture of overlapped sub-band CNN (Fig.~\ref{fig:subband}).}
    \label{table:arch_subband}
    \begin{tabular}{| p{2.3cm} | l | l | l |}
    \hline
    Layer & Band 1 & Band 2 & Band 3 \\ \hline
    Conv 1 &20$\times$8, $K$,&20$\times$8, $K$,&20$\times$8, $K$, \\
    (t{$\times$}f, ch, stride)  &  1$\times$1  & 1$\times$1 & 1$\times$1 \\\hline
    Activation 1 &ReLU&ReLU&ReLU\\\hline
    Dropout 1 &$P$&$P$&$P$\\\hline
    Maxpool &2$\times$2,&2$\times$2,&2$\times$2,\\
    (t{$\times$}f, stride)  &2$\times$2&2$\times$2&2$\times$2\\\hline
    Concat. (axis) &\multicolumn{3}{|c|}{Channel}\\\hline
    Conv 2 &\multicolumn{3}{|c|}{10$\times$4, $K$, 1$\times$1}\\
    (t{$\times$}f, ch, stride) &\multicolumn{3}{|c|}{}\\\hline
    Activation 2 &\multicolumn{3}{|c|}{ReLU}\\\hline
    Dropout 2 &\multicolumn{3}{|c|}{$P$}\\\hline
    Dense (\# outputs)&\multicolumn{3}{|c|}{12}\\
    \hline
    \end{tabular}
    \vspace{-0.8cm}
\end{center}
\end{table}



\section{Experimental Results}


\subsection{Datasets}
We tested the proposed model on Google Speech Commands dataset~\cite{Warden_GoogleSpeechCommand}, which has 35 words in the latest version (v0.02).
We chose two subsets as our testbed for spoken term classification tasks.
For the formulation of subsets, we use the same setup as the ``Audio Recognition" tutorial in the official Tensorflow package~\cite{tf-asr}.
The task is formulated as a twelve-way classification.
For each task, the subset consists of ten keywords, silence, and unknown (i.e. words not belong to the ten selected keywords).
The first task is commands classification, which contains ten keywords as: ``yes", ``no", ``up", ``down", ``left", ``right", ``on", ``off", ``stop", or ``go".
The second task is digits classification, which uses digits zero to nine as ten keywords.
There are 36,923, 4,445, 4,890 samples for commands classification, and  37,390, 4,373, 4,929 samples for digits classification in train, dev, test sets respectively.

\noindent
\textbf{Feature extraction} 
Each utterance is an one second clip with mono audio signals sampled at 16kHz.
The acoustic features used in this work are mel-frequency cepstral coefficients (MFCCs), and the inputs fed to each sub-band sub-network are actually MFCC sub-vectors.
For each one-second clip, we extract 40 dimensional MFCCs from frames of 30 ms duration with shifts of 10 ms.

\subsection{Experimental Setups}
We compare three weight sharing methods for CNN on spoken term classification: (1) full band (Fig.~\ref{fig:fullband}), (2) full band plus non-overlapped sub-band (Fig.~\ref{fig:multiband}), (3) overlapped sub-band (Fig.~\ref{fig:subband}).
To investigate the performance of different model sizes, we experimented with different number of kernels ($K$) in each convolutional block.
Every curve in Fig.~\ref{fig:acc.main} and Fig.~\ref{fig:acc.ablation} consists of data points generated with different values of $K$: \{8, 16, 24, 32, 40, 48, 56, 64\}.
For each model with a specific $K$, we use floating point operations per second (FLOPS) as the measure of model complexity.
Number of FLOPS is measured by \textit{float\_operation} function in the official Tensorflow profiler tool.
We apply the same dropout probability ($P$=0.5) to all the models in this paper.
All models are trained with stochastic gradient descent (SGD) optimizer with a minibatch size of 100.
We train the models with an intial learning rate of 0.001 for 24k iterations, and drop the learning rate to 0.0001 for another 3k iterations.
For the overlapped sub-band models shown in Fig.~\ref{fig:commands.layers}, the bands are \{[0,26), [14,40)\} for 2 bands, \{[0,16), [12,28), [24,40)\} for 3 bands, and \{[0,14), [8,22), [16,30), [26,40)\} for 4 bands, respectively.

The evaluation metric used for spoken term classification in this work is accuracy.
All the accuracies reported in this paper are the average of five random trials to reduce the effect caused by randomness during the training of CNN models.
Error bars in all figures are the sample standard deviation of five trials.
We chose accuracy as the evaluation metric rather than a Detection Error Tradeoff (DET) of false reject and false accept rate for the easiness to compare a large number of trained models in a plot.
It's more succinct to represent five trials using one data point with an error bar compared to plotting tens of DETs in a figure.

\subsection{Results}
Fig.~\ref{fig:commands.main} and Fig.~\ref{fig:digits.main} show the accuracy curves for different weight sharing methods on commands and digits classification.
We have two major observations from these plots.
First, overlapped sub-band outperforms the other two methods in for both cases.
It shows that using overlapped sub-band CNNs to limit weight sharing within a narrow band works better on spoken term classification.
This observation is aligned with the motivation of this work: spatial invariance property of CNN does not fit acoustic applications well, and feature axis should be treated different from time axis.
Interestingly, full band plus non-overlapped sub-band CNN does not work better than the baseline model.
We hypothesize that the full band sub-network and the non-overlapped sub-band sub-network may extract similar features, which causes redundancy in computation.
Second, the computational efficiency brought by overlapped sub-band CNN is more beneficial for small-footprint models (i.e. in the region with lower FLOPS).
If we set the target accuracy as the baseline model (full band) with 500k FLOPS, the proposed sub-band CNN architecture reduces the computation (in terms of FLOPS) by 39.7\% on commands classification, and 49.3\% on digits classification.
Similarly, if the target accuracy is set as the baseline model with 1,000k FLOPS, the reduction is 23.7\% on commands classification, and 50.1\% on digits classification.
Given the same $K$, we found that the decrease in FLOPS of the proposed method comes from the decrease of number of points in the feature map generated by \textit{conv2}.
Feed inputs with less number of points to the final dense layer significantly reduce the required computation.
From the trend of curves shown in Fig.~\ref{fig:acc.main}, we suspect that full band model and overlapped sub-band model may have similar performance when unlimited computational budget is given.

\subsection{Concatenation of Sub-band Features}
\label{sec:concat}
To investigate the effect of different methods to concatenate feature maps from each sub-band, we tested three settings as following:
(1) \textit{concat-c@conv1}: Concatenate along channel axis after the first convolutional layer. This is the \textit{overlapped sub-band CNN} in all other sections throughout this paper.
(2) \textit{concat@conv2}: Each sub-band sub-network has two convolutional layers, and we concatenate all feature maps after the second convolutional layer.  The axis for concatenation does not matter under this setting since the concatenated feature is further fed into a dense layer.
(3) \textit{concat-f@conv1}: Concatenate along feature axis after the first convolutional layer.
As shown in Fig.~\ref{fig:commands.concat}, \textit{concat-c@conv1} outperforms the other two concatenation methods.
By concatenating along channel axis, the receptive field of each point in the concatenated feature map after the first convolutional layer has been tripled in feature axis (from 21$\times$9 to 21$\times$27).
Larger receptive field enables the feature map responds to large enough areas in the input feature map to capture information about acoustic signals spread across different feature dimensions.
Traditionally, larger receptive field is achieved by stacking more convolutional layers, which is less feasible for mobile devices or smart speakers.
The proposed sub-band CNN provides an alternative way to achieve larger receptive field, which is suited for small-footprint model on spoken term classification.

\section{Conclusions}
In this paper, we proposed a sub-band CNN architecture and explored it for spoken term classification.
We compare the proposed sub-band CNNs to full band CNNs and another weight sharing approach on two spoken term classification tasks.
The proposed architecture of sub-band CNNs reduces the computation by 39.7\% on commands classification, and 49.3\% on digits classification, to achieve the same accuracy as the baseline full band CNN with 500k FLOPS.
We found that the computational efficiency brought by sub-band CNN is more beneficial for small-footprint models.
Potential applications for the proposed architecture include on-device speech command recognition, acoustic event detection, etc.

\section{Acknowledgement}
The authors would like to thank Weiran Wang, Krishna Puvvada, and Wei-Ning Hsu for useful discussions.

\bibliographystyle{IEEEtran}

\bibliography{mybib}


\end{document}